\documentstyle[twocolumn,prl,float,aps,epsfig,rotating]{revtex}
\begin{document}

\draft 

\title{Test of isospin symmetry via low
energy $^1$H($\pi^-$,$\pi^o$)$n$ charge exchange}

\author{%
\mbox{ Y.~Jia,$^{1}$} 
\mbox{ T.~P.~Gorringe,$^{1}$} 
\mbox{ M.~D.~Hasinoff,$^{2}$}
\mbox{ M.~A.~Kovash,$^{1}$} 
\mbox{ M.~Ojha,$^{1}$} 
\mbox{ M.M.~Pavan,$^{3}$} 
\mbox{ S.~Tripathi,$^{1}$} 
\mbox{ P.~A.~{\.Z}o{\l}nierczuk,$^{1}$} 
\\
}

\address{%
\mbox{$^1$  University of Kentucky, Lexington, KY 40506}
\mbox{$^2$  University of British Columbia, Vancouver, B.C., Canada V6T 1Z1}
\mbox{$^3$  TRIUMF, 4004 Wesbrook Mall, Vancouver, B.C., Canada V6T 2A3} 
}

\date{\today}
\maketitle

\begin{abstract}
We report measurements of the $\pi^- p \rightarrow \pi^o n$ differential
cross sections at six momenta (104-143~MeV/c) and four 
angles (0-40~deg) by detection of $\gamma$-ray pairs
from $\pi^o \rightarrow \gamma \gamma$ decays using the TRIUMF 
RMC spectrometer. This region exhibits a 
vanishing zero-degree cross section from
destructive interference between s-- and p--waves, 
thus yielding special sensitivity 
to pion-nucleon dynamics and isospin symmetry breaking. 
Our data and previous data 
do not agree,
with important implications for earlier claims of 
large isospin violating effects in low energy pion-nucleon interactions.
\end{abstract}

\pacs{13.75.Gx,25.80.Gn} 

Studies of pion-nucleon dynamics have long played a
central role in nuclear physics.
For example, the $\pi$N system
enables important tests of basic symmetries --
notably isospin symmetry and chiral symmetry --
that are fundamental
to the low energy realization of the strong force.
Interestingly, the possibility of gross violations
of isospin symmetry in $\pi$N processes 
involving neutral pions
was first suggested by Weinberg \cite{We77}.

Of the eight possible $\pi$N$\rightarrow$$\pi$N channels 
there are three experimentally accessible
channels:  $\pi^+$p elastic scattering,
$\pi^-$p elastic scattering and 
$\pi^- p \rightarrow \pi^o n$ charge exchange (CEX). 
Assuming isospin symmetry,
the three hadronic amplitudes 
for the three channels 
are determined by two isospin amplitudes, 
thus implying a triangle relationship 
among the amplitudes $f_{CEX} = ( f_- - f_+ ) / \sqrt{2}$. 
Consequently, an important test 
of isospin symmetry is possible by comparing the
extracted amplitude, $f_{CEX}$, 
from CEX data,
with the predicted amplitude, 
$f^{\prime}_{CEX} \equiv ( f_- - f_+ ) / \sqrt{2}$, 
from elastic data.

Special sensitivity to isospin violation
in $\pi$N interactions is afforded
by pion CEX at forward angles and 100-150~MeV/c. 
At these kinematics
the dominating s- and p-waves interfere destructively
and yield a striking dip in the cross section \cite{Fi86}.
This cancellation of the isospin conserving amplitude
thus amplifies the sensitivity to possible isospin non-conserving amplitudes,
thereby allowing a delicate test of symmetry breaking.

A comparison between the scattering amplitudes predicted from 
elastic data ($f^{\prime}_{CEX}$) and those extracted 
from CEX data ($f_{CEX}$) was 
made by Gibbs {\it et al.}\ \cite{Gi95,Gi04} and 
Matsinos {\it et al.}\ \cite{Ma97,Ma06}. While different
in details, both  studies 
found evidence of an unexpectedly large isospin violation
after accounting for the effects 
of $\pi^{\pm}$p Coulomb interactions
and hadronic mass differences (we note this conclusion 
was questioned in  Refs.\ \cite{Gr04,Ar04}).
In particular, both Gibbs {\it et al.}\ and Matsinos {\it et al.}\
reported a $-0.012 \pm 0.003$~fm discrepancy 
between the forward scattering amplitudes 
that are extracted from the CEX data
and those predicted from the elastic data. 

Subsequently, Piekarewicz \cite{Pi95} claimed 
that a symmetry breaking originating in the u,d-quark mass difference
may explain these results.
In Piekarewicz's calculation the u,d-quark mass difference
induces inequalities among the $\pi^o$p,
$\pi^o$n and $\pi^{\pm}$N couplings 
that are realized as a 
-$0.015$~fm correction
to the triangle relationship.
However Fettes and Meissner \cite{Fe01},
using chiral perturbation theory, 
have concluded quark mass differences
yield much smaller isospin breaking effects  
in the $\pi$N interaction.

While the elastic dataset is extensive and precise, the 
CEX dataset is more limited and less accurate \cite{SAID}.
Most notably, the region
of momenta 100-150~MeV/c
and forward angles is dominated
by a single experiment due 
to Fitzgerald {\it et al.}\ \cite{Fi86} (an additional experiment
by Isenhower {\it et al.}\ \cite{Is99} has reported preliminary data at
112~MeV/c).
The Fitzgerald experiment used
the LAMPF $\pi^o$ spectrometer and measured
cross sections at momenta 100.6-147.1~MeV/c 
and angles $\theta$$<$$25$~degrees.
Unfortunately, 
both the absolute normalization,
obtained from
$^{12}$C($\pi$,$\pi$N)$^{11}$C activation measurements, 
and the background subtraction,
determined from $^{12}$C($\pi^{-}$,$\pi^o$) background measurements,
have been questioned \cite{Is99,Fr97}.

Given the importance of the cross section minimum
as an isospin symmetry test 
we undertook a new measurement of
CEX differential cross sections
at momenta 104-143~MeV/c and angles $0$-$40$~degrees. 
Our experiment was conducted using the RMC
detector on the M9A beamline at the TRIUMF cyclotron. 
The RMC detector is a large acceptance
pair spectrometer that was developed 
for radiative muon capture studies \cite{Wr92}.

The M9A beamline provided a pion flux 
of 1.9-2.5~$\times$10$^{6}$$s^{-1}$
with $e$/$\mu$ contamination of 
28-58\% for momenta 104-143~MeV/c. 
The spot size was about 3$\times$3~cm$^2$ (FWHM)
and the momentum bite was about 6~MeV/c (FWHM).
Note that the primary proton beam comprised
pulses of duration 2-4~ns and separation 43~ns, 
and therefore the arrival of 
$\pi$'s, $\mu$'s and $e$'s were time-separated.

Beam particles were registered 
in two plastic scintillators before traversing 
the target.
The upstream counter (B1) was 6.0~cm 
diameter and 0.3~cm thick and the downstream counter (B2) 
was 6.0~cm diameter and 0.15~cm thick. 
The targets 
were disks of polyethylene (CH$_2$) and carbon (C) 
of diameter 7.0~cm and thickness 1.2~cm.

Neutral pions were identified by detecting the 
two photons from $\pi^o \rightarrow \gamma \gamma$ decay 
using the RMC spectrometer.
The CH$_2$/C targets were located about 40~cm upstream 
of the geometrical center of the RMC spectrometer,
thus permitting the detection of forward-going photon pairs 
from forward angle CEX. 
The photons were detected 
by $\gamma \rightarrow e^+e^-$ conversion
in a cylindrical lead converter 
of thickness 1.7~mm, radius 24.7~cm and length 60.0~cm.
The resulting $e^+$$e^-$ pairs 
were tracked in a low mass cylindrical 
multi-wire chamber and a large volume 
cylindrical drift chamber.
A four-fold segmented, doubled-layered ring 
of veto scintillators (the A/A$^{\prime}$ ring),
located within the Pb radius,
was used to reject backgrounds from charged particles.
A sixteen-fold segmented, single-layered ring 
of trigger scintillators (the D ring),
located outside the drift chamber radius,
was used to identify $e^{\pm}$'s 
from $\gamma$-ray conversions.
A large solenoid provided a 2.206~kG
axial magnetic field 
thus enabling
the momentum analysis of charged particles.

The $\pi^o$ trigger required coincident hits 
in the two beam counters, zero hits
in the A/A$^{\prime}$-ring,
and $\geq$3 hits in the D-ring.
The $\pi^o$ trigger also imposed a requirement of $\geq 3$ cell hits
in drift chamber layer 2 and $\geq 6$ cell clusters 
in drift chamber layers 3+4.
In addition, a beam trigger
was derived by prescaling the B1$\cdot$B2 
coincidences and used to continuously monitor 
the beam properties.
On fulfilling either trigger
the acquisition read out
the chamber hits and times and the scintillator ADCs and TDCs.

A stopping $\pi^-$ beam and liquid H$_2$ target 
were used for the spectrometer calibration
via at-rest $\pi^- p \rightarrow \pi^o n$
using the well-known Panofsky ratio \cite{Sp77}.
The stopping beam had a momentum of 81~MeV/c 
and the hydrogen target was a cylindrical cell of 
16~cm diameter and 15~cm length with 0.25~mm thick Au walls. 

Our data was collected at six mementa with central values
103.8, 112.9, 118.3, 123.8, 134.4 and 142.8~MeV/c.
At each momenta setting we made measurements
with the CH$_2$, C transmission targets and an empty target setup,
and collected all scattering angles simultaneously.

The analysis was organized as follows. 
First, we applied beam, tracking, photon 
and $\pi^o$ cuts to identify the photon pairs 
from $\pi^o \rightarrow \gamma \gamma$ decays. 
Next, for each $\pi^-$ momentum setting and $\pi^o$
angle range, the number of  $^1$H$( \pi^- , \pi^o )$ events
was extracted via a fitting procedure using the 
CH$_2$, C, and empty target momentum spectra.
Finally, from a simulation 
of the in-flight CEX detection efficiency,
and the measurement of the 
at-rest CEX detection efficiency,
we normalized the $^1$H$( \pi^- , \pi^o )$ events
and determined the cross sections.

The event reconstruction first assembled chamber hits 
into candidate tracks, then paired $e^+$-$e^-$ 
tracks into candidate photons, and finally paired photons
into candidate $\pi^o \rightarrow \gamma \gamma$ events.
A beam cut imposed requirements on the minimum amplitudes of the pulse heights 
in the two beam counters.
A tracking cut imposed requirements
on the minimum points and the maximum variances
in the fits to the helical $e^+$, $e^-$ trajectories 
in the tracking chambers.
A photon cut required that
the $e^+$ and $e^-$ tracks converge at the
Pb converter, and a $\pi^o$ cut required 
that the photon-pair converge at the target. 

\begin{figure}
\begin{center}
\epsfig{file=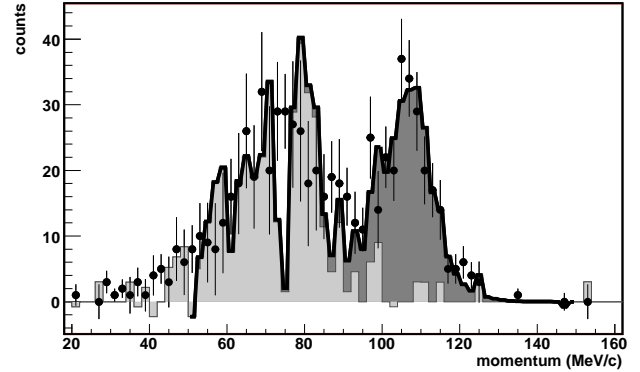,height=5.5cm}
\end{center}
\caption{The $\pi^o$ momentum spectrum for 
an incident $\pi^-$ momentum 104~MeV/c and scattering angles
10-20~deg. The data points (filled circles)
are the measured CH$_2$ spectrum and the solid line is the fit
to the sum of the measured ($\pi^-$,$\pi^o$) background (light gray fill color)
and the $^{1}$H($\pi^-$,$\pi^o$) lineshape (dark gray fill color).}
\label{f: fit}
\end{figure}

The data at each momentum
were divided into four angle bins 
with scattering angles 0-10, 10-20, 20-30 and 30-40~degrees, 
thus yielding twenty-four data points.
Representative $\pi^o$ momenta spectra 
for the CH$_2$ and C targets
are shown in Fig.\ \ref{f: fit}.
The  CH$_2$ spectra show
a peak from the $^1$H$( \pi^-, \pi^o )$ reaction 
and a continuum from the $^{12}$C$( \pi^-, \pi^o )$ reaction.
At forward angles the $^1$H peak and 
$^{12}$C background were kinematically separated while
at larger angles the contributions were overlapping.

To obtain the number of $^1$H$( \pi^-, \pi^o )$ events
we performed fits of the measured CH$_2$ spectra 
to the sum of a simulation of the $^1$H$( \pi^-, \pi^o )$ lineshape
and the measurement of the $( \pi^-, \pi^o )$ background.
Our benchmark fits ({\it e.g.}\ Fig.\ \ref{f: fit}) involved one free parameter,
the normalization constant
for the $^1$H$( \pi^-, \pi^o )$ lineshape.
The background spectra for each angle-momentum bin
was fixed by combining carbon
and empty target spectra in accord
with the target $\pi^-$ exposures and areal densities.

Additional fits were made to quantify
the systematic uncertainties associated with the
$^{1}$H$( \pi^-, \pi^o )$ lineshape, $( \pi^-, \pi^o )$ background, {\it etc}.
In one study we allowed the position
of the $^1$H peak
to be varied in the fit, and found
a mean deviation of the $^1$H yields
from the benchmark fits of 0.51~$\sigma$ (standard deviations).
In a second study we allowed the amplitude
of the background spectra
to be varied in the fit, and found
a mean deviation of the $^1$H yields
from the benchmark fits of 0.37~$\sigma$.
In a third study we compared the results of
fits either including or excluding 
a small correction for the 1.5-2.5~MeV/c differences 
in the CH$_2$-C target energy loss, 
and found a mean deviation of the $^1$H yields of 0.22~$\sigma$.
For the final uncertainties in the $^1$H yields
we took the quadrature sum of the statistical
errors from the benchmark fits with the maximum deviations
from the supplementary fits.

The number of incident pions was determined
by counting the B1$\cdot$B2 coincidences 
and measuring the beam composition. The beam composition
was determined using the beam trigger data 
and was based on the differences in the flight times 
and the pulse heights of the $\pi$'s, $\mu$'s and $e$'s. 
The largest uncertainty in the $\pi^-$ 
flux determination was the beam composition measurement. 

To derive the $^1$H$( \pi^- , \pi^o )$ differential cross sections
we required the spectrometer response function.
The response function 
determines the probability of reconstructing a neutral pion of true
momentum $p_T$ and angle $\theta_T$ 
with measured momentum $p_M$ and angle 
$\theta_M$. It was computed
via a simulation
using the GEANT3 code \cite{GEANT}.
The simulation incorporated
the full geometry of the experimental setup and included 
the detailed interactions of the outgoing photons,
conversion electrons, {\it etc}.
The same analysis package was used to process
both simulated data and measured data.

The absolute normalization of the response function 
was obtained by comparing the
measurement and simulation of at-rest 
CEX for the stopping $\pi^-$ beam 
and the liquid H$_2$ target.
The energy-angle distributions of
$\gamma$-pairs from at-rest CEX
obtained from the measurement and the simulation
were in excellent agreement.
These comparisons were made for different
trigger definitions, cut parameters, {\it etc}.,
in order to fully test the detector simulation.
However, the absolute efficiency for $\pi^o$ detection
was found to be smaller in the measurement than the simulation
by a factor, $F = 0.85 \pm 0.03$. 
This difference is suspected
to arise from contributions such as chamber noise 
and multiple pulsing that were absent in the simulation,
and was observed in previous experiments with the RMC spectrometer.
Importantly, at the level of the quoted uncertainty $\pm$$0.03$, 
the factor $F$ 
was independent of the trigger condition, 
cut parameters, photon energies and target position.

During production running the relative acceptance 
was monitored by measurements of at-rest 
CEX using a 2.5~cm thick CH$_2$ target
and an 81~MeV/c stopping $\pi^-$ beam.
The acceptance was found to be constant to about $\pm$5\% 
over the duration of the experiment.
Given the uncertainties in the determination
of the absolute acceptance and the monitoring 
of the relative acceptance,
an uncertainty of $\pm$10\% was conservatively assigned 
to the overall normalization of the cross sections. 

The measured cross sections 
are averages over the momentum dispersion 
of the beam and the momentum spread 
in the target.
The beam momentum dispersion was obtained 
by analysis of the time-of-flight
spectra of the incident beam particles.
The target momentum spread was obtained 
by a GEANT3 simulation of the beam interactions
in the target material.
The typical beam dispersion was 2.8~MeV/c ($\sigma$)
and the typical target spread was 2.6~MeV/c ($\sigma$).
A correction (of 0.1-0.4~$\sigma$ in magnitude) was applied to
convert the bin-averaged cross sections
to bin-centered cross sections.

\begin{figure}
\begin{center}
\epsfig{file=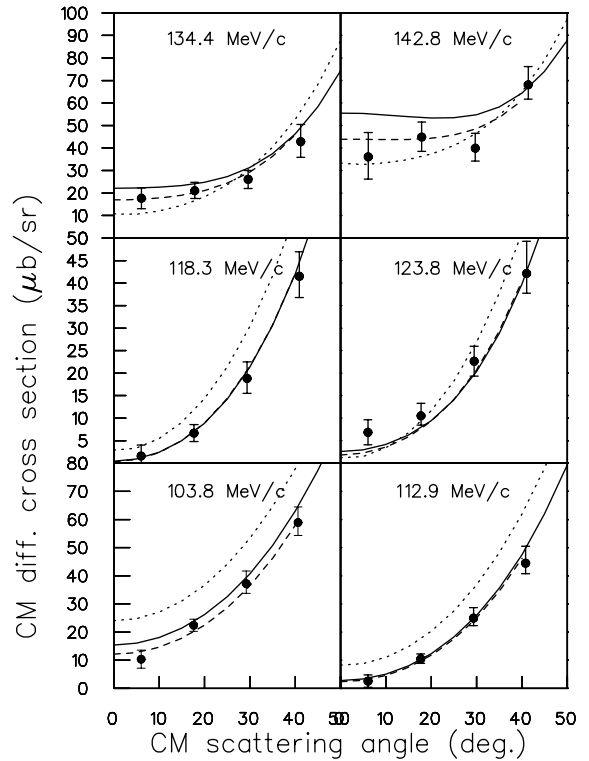,height=10.0cm,angle=0}
\end{center}
\caption{The $\pi^- p \rightarrow \pi^o n$ CM
differential cross section versus central momentum and CM
angle. The data points are our results. The solid line 
is the prediction of Gibbs {\it et al.}\ \protect\cite{Gi95,Gi04}, the dashed line
is the prediction of Matsinos {\it et al.}\ \protect\cite{Ma06}, and the 
dotted line is the prediction of the GWU group \protect\cite{SAID}.}
\label{f: results1}
\end{figure}

Our final results for the center-of-mass (CM) differential cross sections 
versus CM scattering angle and $\pi^-$ laboratory momentum are 
plotted in Fig.\ \ref{f: results1}.
The error bars include the combined uncertainties 
in the $\pi^-$ flux, $^1$H($\pi^-$,$\pi^o$) yields, 
angular resolution and momentum interval, but omit the $\pm$10\% 
normalization uncertainty.

Our data and earlier data near the
minimum are in disagreement.
Compared to Fitzgerald {\it et al.}\  \cite{Fi86} 
our cross sections are roughly factors of two smaller
at momenta $<$120~MeV/c.
This disagreement is beyond the quoted experimental uncertainties 
and -- as discussed in Refs.\ \cite{Is99,Fr97} -- may be related
to the normalization procedure and the background subtraction in the
Fitzgerald {\it et al.}\ experiment.
At 113~MeV/c our results are also $\sim$30\% smaller than the
preliminary results of the Isenhower {\it et al.}\ experiment \cite{Is99},
although this discrepancy is statistically less significant.

Also plotted in Fig.\ \ref{f: results1} are the predicted
cross sections from Gibbs {\it et al.}\ \cite{Gi95,Gi04},
Matsinos {\it et al.}\ \cite{Ma06} and the GWU group \cite{SAID}.
The Gibbs {\it et al.}\ analysis used a potential model
where the parameters of the $\pi$N interaction
were obtained by fitting the $\pi^{\pm}$p elastic dataset only.
The predictions by Matsinos {\it et al.}\ 
and the GWU group were derived from partial wave analyses 
of the $\pi^{\pm}$p elastic dataset. We stress the 
GWU results  in Fig.\ \ref{f: results1} are a `special solution'
based on $T$$<$$100$~MeV elastic data only -- and differ slightly from 
their well-known SP06 solution \cite{SAID}.
All analyses include the isospin violation arising 
from Coulomb effects and hadronic mass splittings.

We first discuss the comparison among the Gibbs {\it et al.}\ and
Matsinos {\it et al.}\ predictions and the CEX data -- 
a comparison \cite{Gi95,Gi04,Ma97,Ma06}
that originally motivated claims of large isospin breaking effects 
in $\pi$N interactions.
Fig.\ \ref{f: results1} shows 
our CEX data and these predictions
are in quite good agreement,
with Matsinos {\it et al.}\ being in agreement at all six momenta and 
Gibbs {\it et al.}\  being in agreement at the four lower momentum values
and differing from our data by about 10-20\% at the two higher momentum values.
In contrast, the original comparisons of these predictions to the Fitzgerald {\it et al.}\  data
showed large discrepancies.
We therefore conclude our data contradict the original evidence for
unexpected isospin violating effects.

However, the GWU results in Fig.\ \ref{f: results1} suggest
some caution is necessary.
Their results differ markedly 
from  Gibbs {\it et al.}\ and Matsinos {\it et al.}\  
in the dip region.
Note these analyses do differ in details; for example
in the omission and the normalizations of certain $\pi^{\pm} p$ datasets.
As mentioned by Matsinos {\it et al.}\ \cite{Ma06}, 
such differences could result in small discrepancies in
the s-and p-wave partial amplitudes that develop into
large discrepancies in the CEX cross section.
Thus final conclusions on isospin breaking 
will likely require a better understanding 
of such discrepancies.

\begin{figure}
\begin{center}
\epsfig{file=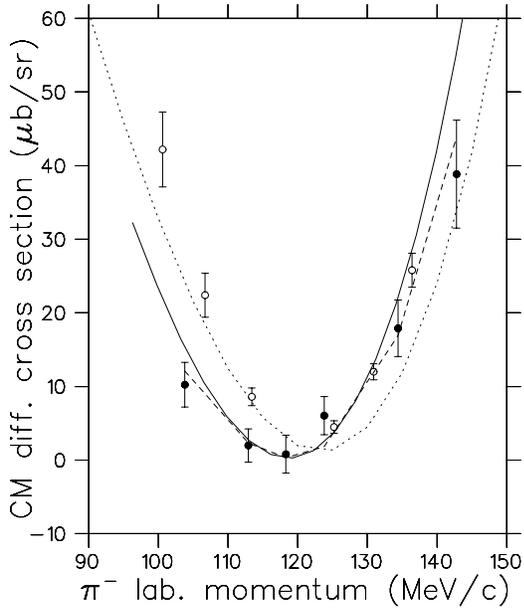,height=8.0cm,angle=0}
\end{center}
\caption{The zero deg.\ $\pi^- p \rightarrow \pi^o n$ CM
differential cross section versus laboratory momentum.
The filled circles are our data,
the open circles are Fitzgerald's data.
The solid line 
is the prediction of Gibbs {\it et al.}\ \protect\cite{Gi95,Gi04}, the dashed line
is the prediction of Matsinos {\it et al.} \protect\cite{Ma06}, and the 
dotted line is the prediction of the GWU group \protect\cite{SAID}.}
\label{f: results2}
\end{figure}

Fig.\ \ref{f: results2} shows the
extrapolated zero-degree cross sections
derived from our data and Fitgerald's data. 
Although the extrapolated cross sections  have larger uncertainties,
they do allow for the direct comparision of the two
experiments.
Like Fitzgerald {\it et al.}, our extrapolations 
have assumed that the cross sections could be parameterized as 
second-order polynomials in $\cos{\theta_{CM}}$.
Fig.\ \ref{f: results2} shows 
our data and Fitzgerald's data are in 
clear disagreement for momenta below 120~MeV/c
but closer agreement for momenta above 120~MeV/c.
Moreover, while our data are reasonably consistent
with Gibbs {\it et al.}\ and Matsinos {\it et al.}\   
at all momenta, the Fitzgerald {\it et al.}\ data are obviously inconsistent with
their predictions below $120$~MeV/c.
The GWU results are in greatest disagreement with our data at momenta $<$$120$~MeV/c
and Fitzgerald's data at momenta $>$$120$~MeV/c.

Last, we performed a fit of our 
zero-degree extrapolations
to a second-order polynomial 
in the $\pi^-$ momentum
and found the minimum to be
$T_{min} = 41.9 \pm 0.9$~MeV.
Our result is in reasonable agreement
with Gibbs {\it et al.}\ and Matsinos {\it et al.}\ 
which yield $T_{min} \simeq 43$~MeV
but in significant disagreement with
the GWU result $T_{min} \simeq 47$~MeV.    
The Fitzgerald {\it et al.}\ value $T_{min} = 45.1 \pm 0.5$~MeV
and our value $T_{min} = 41.9 \pm 0.9$~MeV are in marginal disagreement.

In summary, we have measured the 
CEX differential cross section
at six momenta (104-143~MeV/c) and four angles (0-40~deg.)\
using the TRIUMF RMC spectrometer.
At momenta below 120~MeV/c our cross section measurements 
are considerably smaller than the published data
of Fitzgerald {\it et al.}\ and somewhat smaller than the 
preliminary data of Isenhower {\it et al.}
Interestingly, our results are consistent with the cross sections 
derived from the low energy $\pi^{\pm} p$  elastic dataset 
by Gibbs {\it et al.}\  and Matsinos {\it et al.}, 
and  -- unlike the original comparisons between these analyses and 
earlier data -- show no evidence for unexpected isospin breaking effects.
Further investigations to understand the differences 
between the CEX cross section predictions would be valuable.

We thank Drs.\  R.\ Poutissou and D.\ Healey
for help with the data acquisition and the liquid H$_2$ target, respectively.
We also thank Drs.\ R.\ Arndt, \ W.\ Gibbs, E.\ Matsinos 
and I.\ Strakovsky for 
valuable discussions and the National Science Foundation (United States) 
and the Natural Sciences and Engineering Research Council (Canada)
for their financial support.

\end{document}